\documentclass[a4paper,runningheads]{llncs}

\usepackage[T1]{fontenc}
\usepackage{amsmath,amssymb,amsfonts}
\usepackage{graphicx}
\usepackage{booktabs}
\usepackage{array}
\usepackage{tabularx}
\usepackage{url}
\graphicspath{{figs/}}

\newcommand{\macroeer}{\mathrm{EER}_{\mathrm{macro}}}
\newcommand{\macromindcf}{\mathrm{minDCF}_{\mathrm{macro}}}
\newcolumntype{Y}{>{\raggedright\arraybackslash}X}

\title{OpenSVBench: Diagnosing Speaker Verification
Generalization Across Deployment Scenarios}
\titlerunning{OpenSVBench: SV Generalization Across Deployment Scenarios}

\author{
Qituan Shangguan \and
Bosen Xu \and
Shuai Wang\thanks{Corresponding author.}
}

\authorrunning{Q. Shangguan et al.}

\institute{
School of Intelligence Science and Technology\\
Nanjing University\\
Suzhou, China\\
\email{qituanshangguan@smail.nju.edu.cn}\\
\email{bosenxu@smail.nju.edu.cn}\\
\email{shuaiwang@nju.edu.cn}
}

\begin{document}
\maketitle

\begin{abstract}
Speaker verification (SV) systems now achieve very low error rates on standard public protocols, but those results do not fully describe deployment generalization. VoxCeleb-style evaluations remain useful anchors for English in-the-wild media speech. Scenario-specific evaluations and challenges are also valuable. However, neither setting directly shows whether the same system remains reliable across diverse deployment scenarios.  We introduce OpenSVBench, a public benchmark for cross-scenario SV evaluation.  Its fixed public core contains 26 trial sets organized into 11 scenario groups, with 9.15M verification trials, frozen target/non-target keys, and metadata for scenario-wise reporting.  Around this core, the online leaderboard acts as a diagnostic evaluation interface rather than only a ranked list. It accepts score packages from new systems, evaluates them with frozen keys and common scoring, and reports trial-set breakdowns, scenario-group metrics, macro EER/minDCF, model-wise profiles, and scenario-wise comparisons.  We evaluate 18 public checkpoints spanning x-vector, ECAPA-TDNN, CAMPPlus, ResNet and ERes variants, ReDimNet2, SAM-ResNet, and pretrained Transformer speaker recipes. The results show clear cross-scenario gaps: systems with similar macro EER can have different scenario profiles, strong performance in one scenario does not guarantee strong performance in another, and same-scenario trial sets can still expose source and protocol gaps.

\end{abstract}

\section{Introduction}
\label{sec:intro}

Recent speaker verification systems have advanced rapidly on standard public protocols.  Systems based on TDNN x-vectors \cite{snyder2018xvectors}, ECAPA-TDNN \cite{desplanques2020ecapa}, ResNet-style speaker embeddings \cite{he2016resnet}, and pretrained speech representations \cite{chen2022wavlm,chen2022unispeechsat,chung2021w2vbert} have all improved performance on common benchmarks.  Much of this progress is still measured on VoxCeleb-style evaluation, especially VoxCeleb1-O/E/H \cite{nagrani2020voxceleb}. These protocols are important because they are widely understood and provide a stable anchor for English in-the-wild media speech. Modern systems also report very low error rates on them. For example, a w2v-BERT 2.0 based SV system reports 0.14\% EER on VoxCeleb1-O without score calibration \cite{li2026w2vbert2sv}. Such results show strong progress on a common public anchor, but they leave open how well systems generalize to other deployment conditions.

Real deployments introduce broader enrollment/test mismatch.  Test utterances may be very short, recorded at far-field distance, captured by another device or channel, compressed by a different codec or bandwidth, corrupted by noise or reverberation, or mixed with overlapping speakers.  Speaker traits and speaking style can also vary. Accent, dialect, language, emotion, whisper, Lombard speech, and long time gaps can all change the acoustic evidence available for matching.  A model that performs well on VoxCeleb-style media speech therefore may not generalize to these broader conditions.

Public evaluations have expanded beyond this standard media-speech setting. The NIST SRE series established methodology for speaker recognition evaluation \cite{greenberg2020nist_sre}. CN-Celeb, 3D-Speaker, and FFSVC broaden public evaluation toward non-English media speech, device variation, recording distance, dialect, and channel mismatch \cite{fan2020cnceleb,zheng20233dspeaker,ffsvc2022}. Robustness suites such as SVeritas cover stress tests including natural variation, acoustic corruption, codec compression, demographic factors, spoofing, and adversarial attacks \cite{baali2025sveritas}.  These resources are valuable, but many remain scenario-specific.  They show whether a model is strong for one condition, such as cross-lingual speech, far-field recording, device mismatch, or noisy/reverberant speech. They do not directly provide a unified view of deployment generalization across scenarios.  A cross-lingual evaluation such as TidyVoiceX2-ASV \cite{farhadipour2026tidyvoice,tidyvoice} can identify language transfer behavior, but it does not show whether the same system also generalizes to short-duration, overlap, distance, or device mismatch.

The central question is therefore not only which system is best on a familiar protocol. We also need to know which systems remain reliable across deployment scenarios and where their performance breaks down. OpenSVBench addresses this gap by organizing public SV data into frozen trial sets and scenario groups with metadata. Systems can then be compared by trial-set metrics, scenario profiles, and macro summaries rather than by a single aggregate score. These views are implemented in the online leaderboard, turning the benchmark from a static protocol collection into a shared diagnostic platform for submitting new systems, comparing them with existing models, and analyzing scenario-level failure modes.

This paper makes four contributions.  First, we define OpenSVBench as a benchmark framework for cross-scenario SV generalization.  Second, we construct a fixed public core with 16 sources, 26 trial sets, 11 scenario groups, 9,150,366 verification trials, target/non-target keys, and metadata. Third, we build a diagnostic leaderboard around this fixed core. New systems submit score packages, are evaluated with the same frozen keys and scoring protocol, and are compared through trial-set breakdowns, scenario-group metrics, macro EER/minDCF, model-wise profiles, and scenario-wise comparisons.  Fourth, we analyze 18 released checkpoints with these reporting views. The results show that macro EER alone does not characterize deployment generalization: similar macro scores can hide different failure modes, scenario profiles vary across released recipes and training sources, and some scenario groups contain source/protocol gaps across trial sets.

\section{OpenSVBench Design}
\label{sec:design}

\begin{figure}[tbp]
\centering
\includegraphics[width=\textwidth,trim=14 43 47 76,clip]{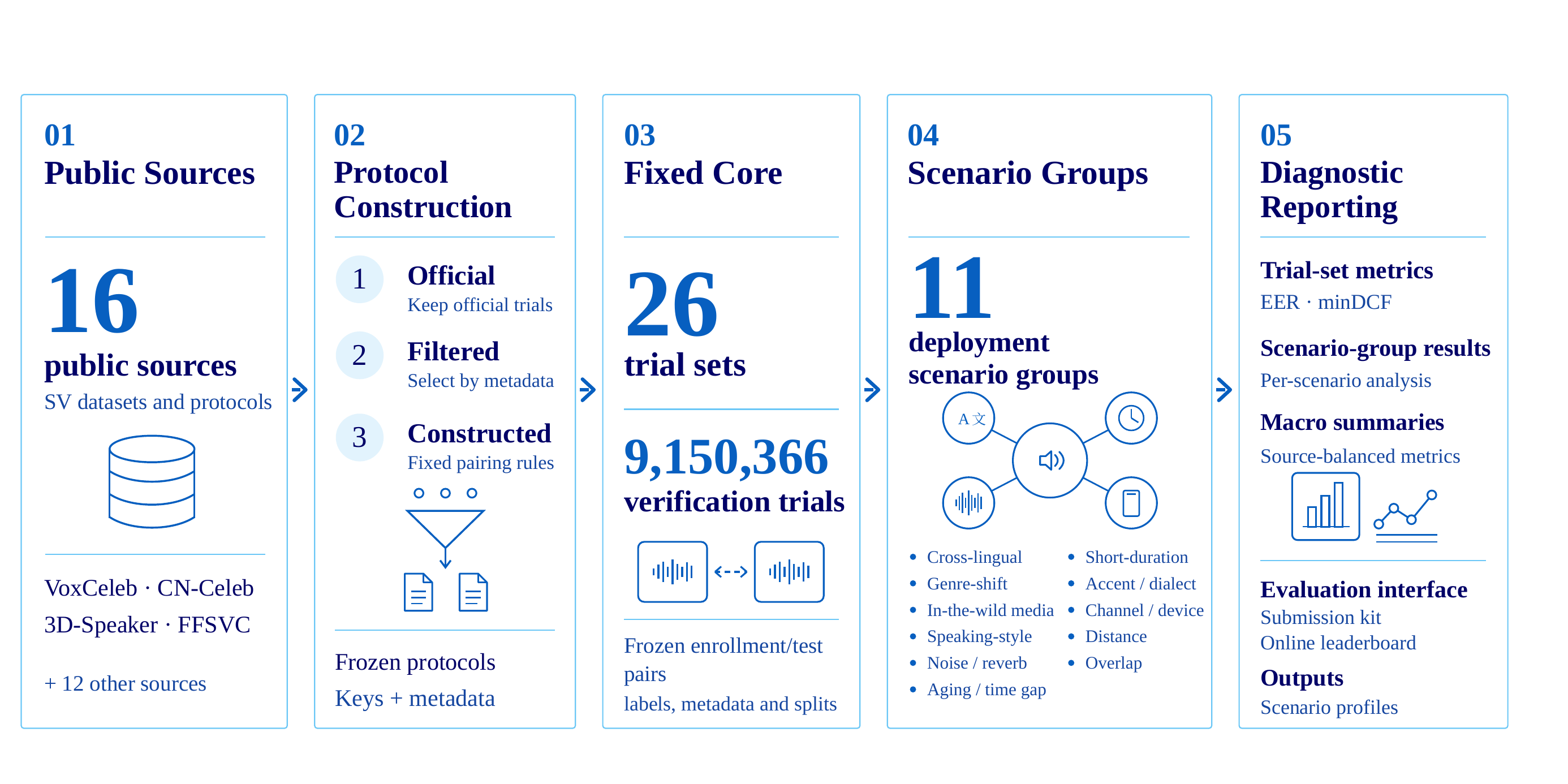}
\caption{Overview of OpenSVBench. Public sources are converted into 26 fixed
verification trial sets across 11 scenario groups using official, filtered, or
constructed protocols. The benchmark reports trial-set, scenario-group, macro,
and EER/minDCF metrics; the submission kit and leaderboard support standardized
score evaluation and diagnostic reporting.}
\label{fig:overview}
\end{figure}

Figure~\ref{fig:overview} summarizes the fixed-core construction and reporting workflow used to evaluate deployment generalization.

\subsection{Design Goals and Principles}
\label{ssec:principles}

OpenSVBench measures cross-scenario SV generalization through cross-scenario coverage, scenario-wise reporting, and standardized public comparison. The fixed core freezes public trial lists, target/non-target keys, metadata, and checksums, while exposing trial-set and scenario-group metrics alongside macro scores. A reviewed contribution workflow supports future scenario additions without changing the fixed-core results reported here.

\begin{table}[t]
\caption{Scenario-level summary of the fixed public core. Counts are aggregated
over trial sets; total verification trials: 9,150,366.}
\label{tab:core}
\centering
\footnotesize
\setlength{\tabcolsep}{2.5pt}
\renewcommand{\arraystretch}{1.05}
\begin{tabularx}{\textwidth}{@{}>{\raggedright\arraybackslash}p{2.2cm}>{\centering\arraybackslash}p{0.7cm}Yrr>{\raggedright\arraybackslash}p{2.3cm}@{}}
\toprule
Scenario & Sets & Sources & Tar & Non & Shift \\
\midrule
Cross-lingual & 1 & TidyVoiceX2 & 17,817 & 182,183 & cross-language \\
Short-duration & 4 & HI-MIA, GSC, CN-Celeb, VoxCeleb & 165,608 & 1,656,080 & short-duration \\
Genre-shift & 1 & CN-Celeb & 40,000 & 400,000 & genre-shift \\
Accent/dialect & 2 & GLOBE, 3D-Speaker & 35,168 & 201,680 & accent/dialect \\
In-the-wild media & 4 & CN-Celeb, VoxCeleb1 & 601,966 & 4,050,649 & media speech \\
Channel/device & 2 & 3D-Speaker, FFSVC & 66,000 & 186,000 & channel/device \\
Speaking-style & 3 & ESD, Whisper40, Lombard & 44,284 & 440,248 & speaking-style \\
Distance & 4 & 3D-Speaker, AliMeeting, CHiME-6, FFSVC & 81,736 & 416,460 & distance \\
Noise/reverb & 1 & VOiCES & 5,000 & 50,000 & noise/reverb \\
Overlap & 2 & AliMeeting, CHiME-6 & 18,600 & 186,000 & overlap \\
Aging & 2 & VoxKnesset, VoxPopuli & 28,325 & 276,562 & aging/time gap \\
\midrule
Total & 26 & All scenario groups & 1,104,504 & 8,045,862 & all trials \\
\bottomrule
\end{tabularx}
\vspace{0.4ex}
\begin{minipage}{\textwidth}
\footnotesize
\raggedright
Detailed construction files and source splits are provided in the benchmark repository and submission kit.
\end{minipage}
\end{table}

\subsection{Trial Sets, Scenario Groups, and Profile Reporting}
\label{ssec:units}

A \emph{trial set} is the scoring unit. It fixes enrollment/test utterances, target/non-target keys, condition labels, metadata, and split information.  A \emph{scenario group} collects one or more trial sets for a deployment condition.  A system's scenario profile is the vector of its scenario group metrics, while trial-set results keep source and protocol details visible.

\subsection{Fixed Core and Protocol Construction}
\label{ssec:sources}

The fixed core is organized around deployment shifts that commonly affect speaker matching. It contains 11 scenario groups: cross-lingual, short-duration, genre-shift, accent/dialect, in-the-wild media speech, channel/device, speaking-style, distance, noise/reverb, overlap, and aging. These groups cover diverse deployment shifts and should be interpreted as generalization views rather than isolated causal factors. Because public sources can entangle language, channel, genre, distance, noise, reverberation, and speaking-style, OpenSVBench keeps the sources and trial-set composition visible for each scenario group.

The fixed core retains widely used public SV protocols for comparability with prior work, including VoxCeleb1-O/E/H, CN-Celeb, and 3D-Speaker views, but it does not treat any one of them as the full evaluation target. Familiar protocols and condition-focused trial sets are organized under the same scenario-group structure, so common anchors and deployment shifts contribute to a unified view of generalization. The full core draws from 16 public sources, including challenge and evaluation corpora \cite{ffsvc2022,richey2018voices,watanabe2020chime6,farhadipour2026tidyvoice,tidyvoice}, large speaker and media-speech corpora \cite{fan2020cnceleb,nagrani2020voxceleb,zheng20233dspeaker,marmor2026voxknesset,wang2021voxpopuli}, and condition-focused sources \cite{hi_mia,warden2018speechcommands,wang2024globe,whisper40,alghamdi2018lombard,zhou2021esd,yu2022alimeeting}. Together they provide 26 ranked trial sets and 9,150,366 verification pairs. Table~\ref{tab:core} summarizes the scenario composition, source coverage, and trial counts.

The core does not require every source to contribute its official test split. Some public sources expose the needed speakers, condition metadata, or usable trials in a development split, training split, evaluation pool, or released challenge development trials.  These choices are fixed in the protocol and recorded in the released construction files.

OpenSVBench uses three construction modes.  Official protocols keep the upstream enrollment and test trial list, target/non-target keys, and evaluation split. This mode is used for VoxCeleb1-O/E/H, CN-Celeb, and the 3D-Speaker device, distance, and dialect trials.  Filtered protocols start from official or public trial lists and retain pairs whose released metadata define the scenario view. In the current core, this covers TidyVoice language views and FFSVC phone/tablet and close-to-far distance views.  When a filtered protocol needs additional non-target pairs, they are added by deterministic rules within the same source and split.  Constructed protocols are used when no suitable upstream SV trial list exists. They form target trials from same-speaker utterance pairs satisfying the scenario condition and non-target trials from different-speaker pairs in the same source and split when possible.

\subsection{Metadata and Condition Labels}
\label{ssec:metadata}

During trial construction, scenario-defining attributes such as device, distance, language, speaking-style, overlap, and aging are selected from official metadata, public annotations, or fixed metadata-derived rules. Other source, session, speaker, and split identifiers are retained to keep unintended variation transparent. These metadata support condition summaries such as duration buckets and overlap severity without changing the ranked core.

\subsection{Submission Kit, Leaderboard, and Public Scenario Additions}
\label{ssec:kit}
The submission kit standardizes evaluation on the fixed public core, covering data preparation, score generation, prechecks, and score package creation. It verifies frozen trial lists and matches scores by utterance-pair keys. All source audio remains subject to the licenses and access terms of the original datasets. The submission kit does not redistribute source recordings; it provides dataset-specific download/access and preparation scripts or local-placement instructions for users to obtain and prepare the required data. The kit is publicly available at \url{https://github.com/shangguanqituan/opensvbench-submission-kit}.

The online leaderboard evaluates and compares submitted systems on the fixed public core and is publicly available at \url{https://opensvbench.xyz/}. It accepts score packages containing submission metadata and per-trial-set score CSV files, then validates them against the frozen trial keys. It computes EER and normalized minDCF with the common scoring protocol and reports the same trial-set, scenario group, and scenario macro views used in this paper. Public ranking is therefore only one view of the benchmark output. The leaderboard also exposes trial-set breakdowns, scenario-level tables, model-wise profiles, and scenario-wise comparisons, helping users identify failure modes under particular deployment shifts.

Reviewed public contribution packages may be considered for later releases, broadening future coverage without changing the fixed-core results reported here.

\subsection{Metrics and Aggregation}
\label{ssec:metrics}

Using the trial-set and scenario-group units defined in Section~\ref{ssec:units}, every trial set is scored separately before aggregation. EER is computed from the submitted raw scores by threshold sweeping.  For a threshold $\theta$, the detection cost is
\begin{equation}
D(\theta) =
C_{\mathrm{miss}}P_{\mathrm{tar}}P_{\mathrm{miss}}(\theta) +
C_{\mathrm{fa}}(1-P_{\mathrm{tar}})P_{\mathrm{fa}}(\theta).
\end{equation}
The normalized minDCF is the minimum of $D(\theta)$ over score thresholds, divided by $\min(C_{\mathrm{miss}}P_{\mathrm{tar}}, C_{\mathrm{fa}}(1-P_{\mathrm{tar}}))$.  We use $P_{\mathrm{tar}}=0.05$ and $C_{\mathrm{miss}}=C_{\mathrm{fa}}=1$ for all submissions.  Because minDCF is minimized over thresholds, it serves here as an operating-point metric under a common scoring protocol.

Scenario group metrics are computed after trial-set scoring.  If scenario group $g$ contains $m_g$ source datasets and source $s$ contributes the trial-set subset $\mathcal{T}_{g,s}$ with trial-set EERs $e_{g,s,i}$, the scenario EER is
\begin{equation}
E_g = \frac{1}{m_g}\sum_{s=1}^{m_g}\left(\frac{1}{|\mathcal{T}_{g,s}|}\sum_{i \in \mathcal{T}_{g,s}} e_{g,s,i}\right).
\end{equation}
The ranked comparison then averages over the $G=11$ scenario groups:
\begin{equation}
\macroeer = \frac{1}{G}\sum_{g=1}^{G} E_g .
\end{equation}
Scenario macro minDCF uses the same source-balanced aggregation rule.  This aggregation prevents very large trial sets and repeated views from one source from dominating the summary, while keeping smaller scenario groups visible.  The macro score provides a compact comparison; interpretation still relies on the scenario profiles and trial-set breakdowns defined above.

\section{Evaluated Systems}
\label{sec:systems}

We evaluate 18 public SV checkpoints released through public model cards or open-source frameworks \cite{wolf2020transformers,wang2023wespeaker,wang2024advancing,ravanelli2021speechbrain,redimnet2}. They cover TDNN/x-vector, ECAPA-TDNN, CAMPPlus, ResNet and ERes variants, ReDimNet2, SAM-ResNet, and pretrained Transformer-based speaker recipes. Table~\ref{tab:overall} lists the system labels and released training sources.

\begin{table}[!t]
\caption{Overall ranking of 18 public systems. Train summarizes released
training sources; macro EER/minDCF use source-balanced scenario aggregation.}
\label{tab:overall}
\centering
\footnotesize
\setlength{\tabcolsep}{3.5pt}
\renewcommand{\arraystretch}{1.04}
\begin{tabularx}{\textwidth}{@{}r p{2.15cm} Y p{1.6cm} r r@{}}
\toprule
Rank & System & Family & Train & $\macroeer$ (\%) & $\macromindcf$ \\
\midrule
1 & W2V-BERT2 & W2V-BERT2 & VC2+VB2 & 9.96 & 0.361 \\
2 & ReDimNet2 & ReDimNet2 & VC2 & 10.53 & 0.419 \\
3 & SAM-R100 & SAM-ResNet100 & VB2+VC2 & 10.58 & 0.390 \\
4 & Res152 & ResNet r-vector & VC2 & 11.23 & 0.443 \\
5 & Res293 & ResNet r-vector & VC2 & 11.31 & 0.440 \\
6 & ERes2V2-ZH & ERes2NetV2 & CN-200k & 11.54 & 0.485 \\
7 & Res34-Vox & ResNet r-vector & VC2 & 12.05 & 0.477 \\
8 & ERes2-EN & ERes2Net & VC2 & 12.10 & 0.473 \\
9 & CAMP-Vox & CAMPPlus & VC2 & 13.42 & 0.521 \\
10 & SB-ECAPA & ECAPA-TDNN & VC1+VC2 & 13.55 & 0.544 \\
11 & ECAPA1024 & ECAPA-TDNN & VC2 & 14.08 & 0.534 \\
12 & ERes2-3D & ERes2Net Large & 3DSpk & 14.17 & 0.594 \\
13 & ECAPA512 & ECAPA-TDNN & VC2 & 14.31 & 0.541 \\
14 & Res34-CN & ResNet r-vector & CN-Celeb & 14.49 & 0.608 \\
15 & IIC-CAMP & CAMPPlus & CN-200k & 15.17 & 0.590 \\
16 & x-vector & x-vector TDNN & VC1+VC2 & 23.35 & 0.776 \\
17 & WavLM-SV & WavLM-SV & SSL+VC1 & 25.16 & 0.875 \\
18 & UniSAT-SV & UniSpeech-SAT-SV & SSL+VC1 & 25.67 & 0.891 \\
\bottomrule
\end{tabularx}

\vspace{2pt}
\begin{minipage}{\textwidth}
\footnotesize
VC1/VC2: VoxCeleb1/2; VB2: VoxBlink2; CN-200k: IIC/ModelScope Chinese
speaker corpus; 3DSpk: 3D-Speaker; SSL+VC1: self-supervised pretraining with
VoxCeleb1 SV fine tuning.
\end{minipage}
\end{table}

All checkpoints are evaluated on the same 26 fixed core trial sets with the same scoring protocol.  The goal is to compare released systems as cross-scenario generalization profiles, not to isolate architecture effects. Released checkpoints differ in training data, augmentation, losses, pooling, adaptation, calibration, and scoring recipes.  Some training sources are also represented in OpenSVBench, such as VoxCeleb, CN-Celeb, or 3D-Speaker; the Train column provides that context.

For the released checkpoint systems, audio is decoded, mixed to mono when needed, resampled to 16 kHz, and passed to the checkpoint without additional fine tuning.  Long utterances can be split into fixed non-overlapping chunks; their embeddings are averaged and L2 normalized. Scores are cosine similarities between L2-normalized enrollment and test embeddings.  We do not use PLDA \cite{ioffe2006plda}, score normalization \cite{auckenthaler2000score_norm}, calibration \cite{brummer2006calibration}, or model-specific backend tuning for these baselines.

\section{Results and Analysis}
\label{sec:results}

\subsection{Overall Ranking and Scenario Profiles}
\label{ssec:profiles}

\begin{figure}[t]
\centering
\includegraphics[width=\textwidth]{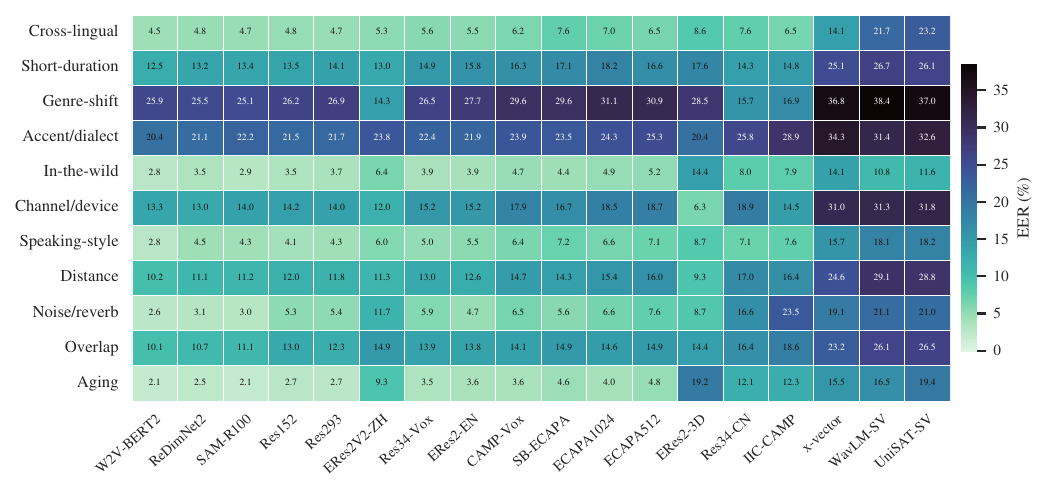}
\caption{Scenario EER profiles for 18 systems. Rows are scenario groups; columns
follow Table~\ref{tab:overall}. Values are source-balanced EER (\%); darker
cells indicate weaker generalization.}
\label{fig:heatmap}
\end{figure}

Table~\ref{tab:overall} gives a compact macro ranking, while Figure~\ref{fig:heatmap} shows the corresponding scenario profiles.  W2V-BERT2, ReDimNet2, SAM-ResNet100, Res152, and Res293 form a strong top group, with macro EER between 9.96\% and 11.31\%.  At the opposite end, x-vector, WavLM-SV, and UniSAT-SV define weaker baselines.  This gap should not be read only as a matter of model age or scale. The two evaluated SSL-to-SV releases show that a strong pretrained speech representation does not automatically yield strong deployment generalization after SV adaptation.

The heatmap shows why the macro ranking is only a compact summary.  Even the strongest systems still struggle on genre-shift and accent/dialect.  For W2V-BERT2, ReDimNet2, SAM-ResNet100, and the deeper WeSpeaker ResNets, genre-shift remains around 26--27\% EER and accent/dialect remains around 20--22\%.  At the same time, ERes2-3D ranks only twelfth by macro EER but is relatively strong on channel/device and distance.  These profile differences show that macro rank is useful for ordering systems, but it cannot replace scenario-wise diagnostics when the goal is deployment generalization.

\subsection{High-Error Scenario Groups in the Fixed Core}
\label{ssec:difficulty}

\begin{figure}[t]
\centering
\includegraphics[width=\textwidth]{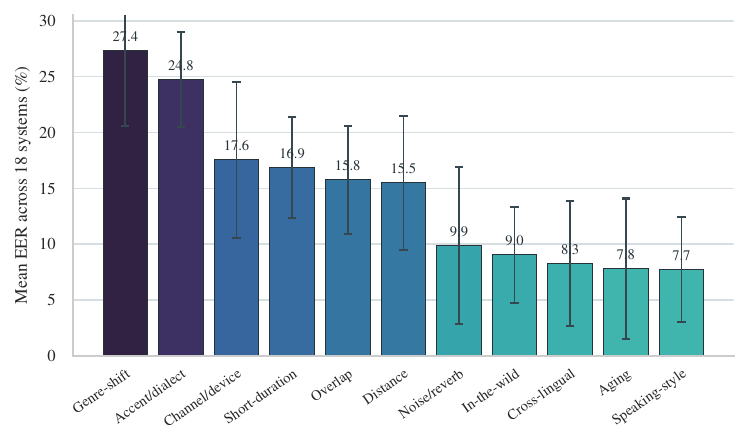}
\caption{Source-balanced scenario EER averaged over 18 systems. Error bars show
standard deviation; darker bars indicate higher mean EER.}
\label{fig:difficulty}
\end{figure}

Figure~\ref{fig:difficulty} reports the mean source-balanced scenario EER of each scenario group across the 18 evaluated systems.  Genre-shift has the highest mean EER at 27.36\%, followed by accent/dialect at 24.76\%.  These are the two hardest groups in the current fixed core, suggesting that public checkpoints remain especially vulnerable to content-domain shifts and to speaker population or trait variation.  Channel/device, short-duration, overlap, and distance form a second high-error cluster, with mean EER between 15.49\% and 17.58\%. These groups correspond to common deployment mismatches that remain far from solved.

The lower-mean groups in Figure~\ref{fig:difficulty} should not be read as fully solved conditions.  In the current fixed core, some groups are narrower than others or rely on a single trial set, so their average difficulty reflects both the intended mismatch and the source protocol.  These scenario means therefore need to be read together with Table~\ref{tab:core} and the trial-set breakdowns below.

\subsection{Condition-Level Diagnostics Explain Where Generalization Breaks}
\label{ssec:conditions}

Condition metadata help explain which subconditions drive the group-level gaps.  The values below are condition-bin mean EERs over the 18 evaluated systems within one concrete source view. They should therefore be read as diagnostics of failure drivers rather than as additional ranking metrics.

Short-duration provides a clear example.  In the CN-Celeb Short view, mean EER falls from 36.1\% at 0.5 s to 15.6\% at 3 s.  In the VoxCeleb Short view, it falls from 31.2\% at 0.5 s to 4.7\% at 3 s.  Both sources improve as duration increases, but the absolute error level and the magnitude of improvement differ substantially.  The diagnostic point is therefore not only that short speech is hard. The duration effect is also source-dependent in the current benchmark.

Overlap severity shows the same pattern.  In AliMeeting, light, medium, and heavy overlap yield mean EERs of 4.8\%, 7.9\%, and 17.6\%.  In CHiME-6, the corresponding values are 14.8\%, 15.6\%, and 30.0\%.  Both sources become harder as overlap increases, but they start from different baselines and separate more strongly at higher severity.  Even under the same condition label, recording setup and protocol still matter. These metadata are therefore explanatory: they identify failure drivers behind one scenario average.

\subsection{Within-Scenario Trial Consistency and Cross-Trial Generalization}
\label{ssec:trial-consistency}

We next ask whether system rankings agree across trial sets inside the same multi-trial scenario group.  For each trial-set pair, we compute Spearman rank correlation over the 18 systems using trial-set EER.  This correlation measures ranking consistency, not absolute difficulty similarity.  Table~\ref{tab:trial_consistency} summarizes these pairwise correlations.  High values indicate similar system ordering within one scenario group.  Low values indicate source, protocol, or domain gaps inside the group. They should be read as diagnostic evidence rather than as a failure of the grouping itself.

\begin{table}[t]
\caption{Within-scenario rank consistency. Pairwise Spearman $\rho$ is computed
over EER rankings across the 18 systems.}
\label{tab:trial_consistency}
\centering
\footnotesize
\setlength{\tabcolsep}{2.5pt}
\renewcommand{\arraystretch}{1.05}
\begin{tabularx}{\textwidth}{@{}>{\raggedright\arraybackslash}p{2.5cm}>{\centering\arraybackslash}p{0.7cm}>{\centering\arraybackslash}p{1.15cm}>{\centering\arraybackslash}p{1.25cm}Y@{}}
\toprule
Scenario & Sets & Mean $\rho$ & Range & Lowest-correlation pair \\
\midrule
Short-duration & 4 & 0.51 & 0.22--0.88 & CN-Celeb Short vs. VoxCeleb Short, $\rho=0.22$ \\
Accent/dialect & 2 & 0.55 & 0.55 & GLOBE vs. 3D-Speaker Dialect, $\rho=0.55$ \\
In-the-wild & 4 & 0.56 & 0.13--0.99 & CN-Celeb vs. VoxCeleb1-E, $\rho=0.13$ \\
Channel/device & 2 & 0.81 & 0.81 & 3D-Speaker Device vs. FFSVC 2022 Cross-Channel, $\rho=0.81$ \\
Speaking-style & 3 & 0.73 & 0.57--0.97 & Whisper40 Whisper vs. Lombard Grid Lombard, $\rho=0.57$ \\
Distance & 4 & 0.64 & 0.30--0.89 & CHiME-6 Far-field vs. FFSVC Cross-Domain, $\rho=0.30$ \\
Overlap & 2 & 0.43 & 0.43 & AliMeeting Overlap vs. CHiME-6 Overlap, $\rho=0.43$ \\
Aging & 2 & 0.94 & 0.94 & VoxKnesset vs. VoxPopuli Aging, $\rho=0.94$ \\
\bottomrule
\end{tabularx}
\end{table}

Aging and channel/device show high agreement, so their trial sets induce relatively consistent system ordering.  Speaking style is also fairly consistent, although not as tightly aligned as aging.  By contrast, short-duration, in-the-wild, and overlap contain low-correlation pairs. A system that ranks well on one source view therefore may not rank similarly on another source view under the same scenario label.  Distance lies between these extremes: it shows a shared scenario trend, but still contains meaningful source and protocol gaps.  This supports OpenSVBench's two-level reporting design. Scenario metrics summarize deployment shifts, while trial-set breakdowns expose rank reversals that the group average cannot show.

\subsection{What Explains Profile Differences Beyond the Macro Score?}
\label{ssec:overall}

The comparisons in this section are not controlled ablations.  Each released checkpoint combines an architecture with a particular training source, augmentation strategy, loss, pooling method, backend, and scoring recipe.  The goal is therefore not to assign causal credit to one factor. Instead, we use the diagnostic views of the benchmark to identify factors associated with the scenario profile differences in Table~\ref{tab:overall} and Figure~\ref{fig:heatmap}.

Source overlap is one useful diagnostic variable, but it is not a sufficient explanation of deployment generalization.  Here source overlap means that a released training description explicitly names a public benchmark source that also appears in OpenSVBench, such as VoxCeleb, CN-Celeb, or 3D-Speaker.  For each eligible model and scenario group, we first average repeated trial-set views within each source. We then compare the overlapping source with the other sources in the same scenario group.

The expected same-source advantage is visible in the easiest case. Across the 14 checkpoints whose released training descriptions include VoxCeleb, the in-the-wild group yields mean EERs of 2.24\% on the VoxCeleb side and 16.09\% on the CN-Celeb side.  However, this signal weakens sharply once the view changes.  In short-duration, the same 14 models still favor VoxCeleb Short in 13 of 14 cases, but the across-model means are much closer: 16.06\% on VoxCeleb Short versus 17.85\% on CN-Celeb Short, GSC Short, and HI-MIA.  Res34-CN is a counterexample in the other direction. Although its released training source names CN-Celeb, it is worse on CN-Celeb Short than on the mean of the other short-duration sources, 17.67\% versus 13.16\%. It is still slightly better on CN-Celeb than on VoxCeleb in the in-the-wild group, 7.47\% versus 8.21\%.  ERes2-3D provides a within-source counterexample. It is much better on 3D-Speaker Dialect than on GLOBE, 12.0\% versus 28.89\%, but this advantage does not extend to the 3D-Speaker channel/device or distance views. There, the 3D-Speaker side is not better than the mean of the other sources in the same group.  Source overlap is therefore visible, but it does not fully explain deployment generalization. Transfer within one scenario group, and even within one source, can remain view-specific.  These comparisons are descriptive diagnostics rather than causal evidence of memorization or overfitting, because source difficulty and protocol design also differ.

Same-family comparisons reinforce this point.  The issue is not simply that architectures differ. Macro scores and architecture labels are both too coarse to explain scenario profiles.  CAMP-Vox and IIC-CAMP share the CAMPPlus lineage, yet their profile trade-off is much sharper than their macro gap suggests. IIC-CAMP is much better on genre-shift (16.9\% vs. 29.6\%) and channel/device (14.5\% vs. 17.9\%), while CAMP-Vox is better on accent/dialect (23.9\% vs. 28.9\%), noise/reverb (6.5\% vs. 23.5\%), overlap (14.1\% vs. 18.6\%), and aging (3.6\% vs. 12.3\%).  The WeSpeaker ResNet family makes the same point from another angle. Res34, Res152, and Res293 obtain 12.05\%, 11.23\%, and 11.31\% macro EER, but depth is not a monotonic explanation of generalization. Res152 and Res293 trade small advantages across channel/device, overlap, distance, and in-the-wild media speech.  ECAPA variants show that embedding size or implementation label is also not enough. ECAPA512 and ECAPA1024 are close at 14.31\% and 14.08\%, and SB-ECAPA reaches 13.55\%. Their profiles still reverse across genre-shift, accent/dialect, channel/device, speaking style, and distance.  Finally, the contrast between W2V-BERT2 at 9.96\% and WavLM-SV and UniSAT-SV at 25.16\% and 25.67\% shows that the label ``pretrained speech representation'' is too broad. SV adaptation and the released recipe remain critical to deployment generalization.

Two stability checks help interpret these profile differences together with the leaderboard ranking.  First, scenario bootstrap asks whether the top ranking depends on a small number of scenario groups.  Over 10,000 bootstrap resamples of the 11 scenario groups, W2V-BERT2 remains first in 84.3\% of resamples and remains in the top three in 99.4\%. ReDimNet2 and SAM-ResNet100 remain in the top three in 87.8\% and 83.0\% of resamples.  This indicates that the top tier is fairly stable. It also shows that W2V-BERT2 is not absolutely first under every possible scenario reweighting, so rank still depends on scenario composition.  Second, source-removal checks ask whether the main conclusions are driven mainly by source overlap.  After removing VoxCeleb trial sets, W2V-BERT2, ReDimNet2, and SAM-ResNet100 change from 9.96\%, 10.53\%, and 10.58\% to 10.44\%, 11.14\%, and 11.08\%.  Removing CN-Celeb changes Res34-CN from 14.49\% to 14.30\%, and removing 3D-Speaker changes ERes2-3D from 14.17\% to 14.85\%.  These changes are not negligible, but they are modest relative to the broader profile differences discussed above.  The ranking is therefore stable enough to be useful. At the same time, diagnostic scenario and trial-set views remain necessary because system profiles and rank confidence still vary with scenario composition.

\subsection{Limitations}
\label{ssec:interpretation}
The current fixed core has three main limitations. First, the scenario groups are not causal isolations: public sources can entangle language, genre, channel, distance, noise, reverberation, speaking style, and collection protocol. Second, coverage is uneven across scenarios, and some groups currently rely on a single public source or trial set. Later reviewed public additions can broaden scenario coverage. Third, complete training manifests are not always available for released checkpoints, so source overlap can be documented when known but cannot always be fully controlled.

\section{Conclusion}
\label{sec:conclusion}

We introduced OpenSVBench, a public benchmark for evaluating SV generalization across deployment scenarios. Its fixed core contains 16 public sources, 26 trial sets, 11 scenario groups, and 9,150,366 verification trials with frozen keys, metadata, and common scoring. Experiments on 18 public checkpoints show substantial scenario-dependent, recipe-dependent, and trial/source-dependent generalization gaps. Similar macro EER can hide different failure modes, and high-error groups include genre-shift, accent/dialect, channel/device, short-duration, overlap, and distance. The submission kit and diagnostic leaderboard support standardized evaluation, model comparison, scenario-wise analysis, and trial-set breakdowns. Future reviewed additions can broaden coverage without changing the fixed-core results.

\subsubsection*{Acknowledgments}

This work is supported by the China NSFC project (No. 62401377) and Yangtze River Delta Science and Technology Innovation Community Joint Research Project (No. 2024CSJGG1100).

\subsubsection*{Disclosure of Interests}

The authors have no competing interests to declare that are relevant to the content of this article.

\bibliographystyle{splncs04}
\bibliography{refs}

@article{greenberg2020nist_sre,
  author  = {Craig S. Greenberg and Seyed Omid Sadjadi and Lisa Mason and Douglas A. Reynolds},
  title   = {Two Decades of Speaker Recognition Evaluation at the National Institute of Standards and Technology},
  journal = {Computer Speech \& Language},
  volume  = {60},
  pages   = {101032},
  year    = {2020},
  doi     = {10.1016/j.csl.2019.101032}
}

@article{nagrani2020voxceleb,
  author  = {Arsha Nagrani and Joon Son Chung and Weidi Xie and Andrew Zisserman},
  title   = {{VoxCeleb}: Large-Scale Speaker Verification in the Wild},
  journal = {Computer Speech \& Language},
  volume  = {60},
  pages   = {101027},
  year    = {2020},
  doi     = {10.1016/j.csl.2019.101027}
}

@article{zheng20233dspeaker,
  author  = {Siqi Zheng and Luyao Cheng and Yafeng Chen and Hui Wang and Qian Chen},
  title   = {{3D-Speaker}: A Large-Scale Multi-Device, Multi-Distance, and Multi-Dialect Corpus for Speech Representation Disentanglement},
  journal = {arXiv preprint arXiv:2306.15354},
  year    = {2023}
}

@inproceedings{baali2025sveritas,
  author    = {Massa Baali and Sarthak Bisht and Francisco Teixeira and Kateryna Shapovalenko and Rita Singh and Bhiksha Raj},
  title     = {{SVeritas}: Benchmark for Robust Speaker Verification under Diverse Conditions},
  booktitle = {Findings of the Association for Computational Linguistics: EMNLP 2025},
  pages     = {9714--9731},
  address   = {Suzhou, China},
  publisher = {Association for Computational Linguistics},
  year      = {2025},
  doi       = {10.18653/v1/2025.findings-emnlp.516}
}

@inproceedings{snyder2018xvectors,
  author    = {David Snyder and Daniel Garcia-Romero and Gregory Sell and Daniel Povey and Sanjeev Khudanpur},
  title     = {{X}-Vectors: Robust {DNN} Embeddings for Speaker Recognition},
  booktitle = {Proc. ICASSP},
  pages     = {5329--5333},
  year      = {2018},
  doi       = {10.1109/ICASSP.2018.8461375}
}

@inproceedings{desplanques2020ecapa,
  author    = {Brecht Desplanques and Jenthe Thienpondt and Kris Demuynck},
  title     = {{ECAPA-TDNN}: Emphasized Channel Attention, Propagation and Aggregation in {TDNN} Based Speaker Verification},
  booktitle = {Proc. Interspeech},
  pages     = {3830--3834},
  year      = {2020},
  doi       = {10.21437/Interspeech.2020-2650}
}

@inproceedings{he2016resnet,
  author    = {Kaiming He and Xiangyu Zhang and Shaoqing Ren and Jian Sun},
  title     = {Deep Residual Learning for Image Recognition},
  booktitle = {Proc. CVPR},
  pages     = {770--778},
  year      = {2016},
  doi       = {10.1109/CVPR.2016.90}
}

@inproceedings{fan2020cnceleb,
  author    = {Yue Fan and Jiawen Kang and Lantian Li and Kaicheng Li and Haolin Chen and Sitong Cheng and Pengyuan Zhang and Ziyao Zhou and Yunqi Cai and Dong Wang},
  title     = {{CN-Celeb}: A Challenging Chinese Speaker Recognition Dataset},
  booktitle = {Proc. ICASSP},
  pages     = {7604--7608},
  year      = {2020},
  doi       = {10.1109/ICASSP40776.2020.9054695}
}

@article{chen2022wavlm,
  author  = {Sanyuan Chen and Chengyi Wang and Zhengyang Chen and Yu Wu and Shujie Liu and Zhuo Chen and Jinyu Li and Naoyuki Kanda and Takuya Yoshioka and Xiong Xiao and Jian Wu and Long Zhou and Shuo Ren and Yanmin Qian and Yao Qian and Jian Wu and Michael Zeng and Furu Wei},
  title   = {{WavLM}: Large-Scale Self-Supervised Pre-Training for Full Stack Speech Processing},
  journal = {IEEE Journal of Selected Topics in Signal Processing},
  volume  = {16},
  number  = {6},
  pages   = {1505--1518},
  year    = {2022},
  doi     = {10.1109/JSTSP.2022.3188113}
}

@inproceedings{chen2022unispeechsat,
  author    = {Sanyuan Chen and Yu Wu and Chengyi Wang and Shujie Liu and Daniel Tompkins and Zhuo Chen and Furu Wei},
  title     = {{UniSpeech-SAT}: Universal Speech Representation Learning with Speaker Aware Pre-Training},
  booktitle = {Proc. ICASSP},
  pages     = {6152--6156},
  year      = {2022},
  doi       = {10.1109/ICASSP43922.2022.9746314}
}

@inproceedings{chung2021w2vbert,
  author    = {Yu-An Chung and Yu Zhang and Wei Han and Chung-Cheng Chiu and James Qin and Ruoming Pang and Yonghui Wu},
  title     = {{w2v-BERT}: Combining Contrastive Learning and Masked Language Modeling for Self-Supervised Speech Pre-Training},
  booktitle = {Proc. ASRU},
  pages     = {244--250},
  year      = {2021},
  doi       = {10.1109/ASRU51503.2021.9688253}
}

@inproceedings{li2026w2vbert2sv,
  author    = {Ze Li and Ming Cheng and Ming Li},
  title     = {Enhancing Speaker Verification with {w2v-BERT} 2.0 and Knowledge Distillation Guided Structured Pruning},
  booktitle = {Proc. ICASSP},
  pages     = {16462--16466},
  year      = {2026},
  doi       = {10.1109/ICASSP55912.2026.11461727}
}

@article{farhadipour2026tidyvoice,
  author        = {Aref Farhadipour and Jan Marquenie and Srikanth Madikeri and Eleanor Chodroff},
  title         = {{TidyVoice}: A Curated Multilingual Dataset for Speaker Verification Derived from Common Voice},
  journal       = {arXiv preprint arXiv:2601.16358},
  year          = {2026},
  eprint        = {2601.16358},
  archivePrefix = {arXiv},
  note          = {Accepted at ICASSP 2026}
}

@article{tidyvoice,
  author        = {Aref Farhadipour and Jan Marquenie and Srikanth Madikeri and Teodora Vukovic and Volker Dellwo and Kathy Reid and Francis M. Tyers and Ingo Siegert and Eleanor Chodroff},
  title         = {{TidyVoice} 2026 Challenge Evaluation Plan},
  journal       = {arXiv preprint arXiv:2601.21960},
  year          = {2026},
  eprint        = {2601.21960},
  archivePrefix = {arXiv}
}

@inproceedings{hi_mia,
  author    = {Xiaoyi Qin and Hui Bu and Ming Li},
  title     = {{HI-MIA}: A Far-Field Text-Dependent Speaker Verification Database and the Baselines},
  booktitle = {Proc. ICASSP},
  pages     = {7609--7613},
  year      = {2020},
  doi       = {10.1109/ICASSP40776.2020.9053194}
}

@article{warden2018speechcommands,
  author        = {Pete Warden},
  title         = {Speech Commands: A Dataset for Limited-Vocabulary Speech Recognition},
  journal       = {arXiv preprint arXiv:1804.03209},
  year          = {2018},
  eprint        = {1804.03209},
  archivePrefix = {arXiv}
}

@inproceedings{wang2024globe,
  author    = {Wenbin Wang and Yang Song and Sanjay Jha},
  title     = {{GLOBE}: A High-Quality English Corpus with Global Accents for Zero-Shot Speaker Adaptive Text-to-Speech},
  booktitle = {Proc. Interspeech},
  pages     = {1365--1369},
  year      = {2024},
  doi       = {10.21437/Interspeech.2024-70}
}

@inproceedings{ffsvc2022,
  author    = {Xiaoyi Qin and Ming Li and Hui Bu and Shrikanth Narayanan and Haizhou Li},
  title     = {The 2022 Far-Field Speaker Verification Challenge: Exploring Domain Mismatch and Semi-Supervised Learning under the Far-Field Scenario},
  booktitle = {Proc. FFSVC},
  pages     = {1--6},
  year      = {2022},
  doi       = {10.21437/FFSVC.2022-1}
}

@article{whisper40,
  author  = {Jingwen Yang and Ruohua Zhou},
  title   = {{Whisper40}: A Multi-Person Chinese Whisper Speaker Recognition Dataset Containing Same-Text Neutral Speech},
  journal = {Information},
  volume  = {15},
  number  = {4},
  pages   = {184},
  year    = {2024},
  doi     = {10.3390/info15040184}
}

@article{alghamdi2018lombard,
  author  = {Najwa Alghamdi and Steve Maddock and Ricard Marxer and Jon Barker and Guy J. Brown},
  title   = {A Corpus of Audio-Visual Lombard Speech with Frontal and Profile Views},
  journal = {The Journal of the Acoustical Society of America},
  volume  = {143},
  number  = {6},
  pages   = {EL523--EL529},
  year    = {2018},
  doi     = {10.1121/1.5042758}
}

@inproceedings{richey2018voices,
  author    = {Colleen Richey and Maria A. Barrios and Zeb Armstrong and Chris Bartels and Horacio Franco and Martin Graciarena and Aaron Lawson and Mahesh Kumar Nandwana and Mitchell McLaren and Julien van Hout and et al.},
  title     = {Voices Obscured in Complex Environmental Settings ({VOiCES}) Corpus},
  booktitle = {Proc. Interspeech},
  pages     = {1566--1570},
  year      = {2018},
  doi       = {10.21437/Interspeech.2018-2144}
}

@inproceedings{watanabe2020chime6,
  author    = {Shinji Watanabe and Michael Mandel and Jon Barker and Emmanuel Vincent and Ashish Arora and Xuankai Chang and Sanjeev Khudanpur and Vimal Manohar and Daniel Povey and Desh Raj and David Snyder and Aswin Shanmugam Subramanian and Jan Trmal and Bar Ben Yair and Christoph Boeddeker and Zhaoheng Ni and Yusuke Fujita and Shota Horiguchi and Naoyuki Kanda and Takuya Yoshioka and Neville Ryant},
  title     = {{CHiME-6} Challenge: Tackling Multispeaker Speech Recognition for Unsegmented Recordings},
  booktitle = {Proc. CHiME},
  pages     = {1--7},
  year      = {2020},
  doi       = {10.21437/CHiME.2020-1}
}

@inproceedings{zhou2021esd,
  author    = {Kun Zhou and Berrak Sisman and Rui Liu and Haizhou Li},
  title     = {Seen and Unseen Emotional Style Transfer for Voice Conversion with a New Emotional Speech Dataset},
  booktitle = {Proc. ICASSP},
  pages     = {920--924},
  year      = {2021},
  doi       = {10.1109/ICASSP39728.2021.9413923}
}

@inproceedings{yu2022alimeeting,
  author    = {Fan Yu and Shiliang Zhang and Yihui Fu and Lei Xie and Siqi Zheng and Zhihao Du and Weilong Huang and Pengcheng Guo and Zhijie Yan and Bin Ma and Xin Xu and Hui Bu},
  title     = {{M2MeT}: The {ICASSP} 2022 Multi-Channel Multi-Party Meeting Transcription Challenge},
  booktitle = {Proc. ICASSP},
  pages     = {6167--6171},
  year      = {2022},
  doi       = {10.1109/ICASSP43922.2022.9747026}
}

@article{marmor2026voxknesset,
  author        = {Yanir Marmor and Arad Zulti and David Krongauz and Adam Gabet and Yoad Snapir and Yair Lifshitz and Eran Segal},
  title         = {{VoxKnesset}: A Large-Scale Longitudinal Hebrew Speech Dataset for Aging Speaker Modeling},
  journal       = {arXiv preprint arXiv:2603.01270},
  year          = {2026},
  eprint        = {2603.01270},
  archivePrefix = {arXiv}
}

@inproceedings{wang2021voxpopuli,
  author    = {Changhan Wang and Morgane Rivi{\`e}re and Ann Lee and Anne Wu and Chaitanya Talnikar and Daniel Haziza and Mary Williamson and Juan Pino and Emmanuel Dupoux},
  title     = {{VoxPopuli}: A Large-Scale Multilingual Speech Corpus for Representation Learning, Semi-Supervised Learning and Interpretation},
  booktitle = {Proc. ACL},
  pages     = {993--1003},
  year      = {2021},
  doi       = {10.18653/v1/2021.acl-long.80}
}

@inproceedings{wang2023wespeaker,
  author    = {Hongji Wang and Chengdong Liang and Shuai Wang and Zhengyang Chen and Binbin Zhang and Xu Xiang and Yanlei Deng and Yanmin Qian},
  title     = {{WeSpeaker}: A Research and Production Oriented Speaker Embedding Learning Toolkit},
  booktitle = {Proc. ICASSP},
  pages     = {1--5},
  year      = {2023}
}

@article{wang2024advancing,
  author        = {Shuai Wang and Zhengyang Chen and Bing Han and Hongji Wang and Chengdong Liang and Binbin Zhang and Xu Xiang and Wen Ding and Johan Rohdin and Anna Silnova and others},
  title         = {Advancing Speaker Embedding Learning: {WeSpeaker} Toolkit for Research and Production},
  journal       = {Speech Communication},
  volume        = {162},
  pages         = {103104},
  year          = {2024},
  doi           = {10.1016/j.specom.2024.103104}
}

@inproceedings{wolf2020transformers,
  author    = {Thomas Wolf and Lysandre Debut and Victor Sanh and Julien Chaumond and Clement Delangue and Anthony Moi and Pierric Cistac and Tim Rault and R{\'e}mi Louf and Morgan Funtowicz and Joe Davison and Sam Shleifer and Patrick von Platen and Clara Ma and Yacine Jernite and Julien Plu and Canwen Xu and Teven Le Scao and Sylvain Gugger and Mariama Drame and Quentin Lhoest and Alexander M. Rush},
  title     = {Transformers: State-of-the-Art Natural Language Processing},
  booktitle = {Proc. EMNLP System Demonstrations},
  pages     = {38--45},
  year      = {2020},
  doi       = {10.18653/v1/2020.emnlp-demos.6}
}

@article{ravanelli2021speechbrain,
  author  = {Mirco Ravanelli and Titouan Parcollet and Peter Plantinga and Aku Rouhe and Samuele Cornell and Loren Lugosch and Cem Subakan and Nauman Dawalatabad and Abdelwahab Heba and Jianyuan Zhong and Ju-Chieh Chou and Sung-Lin Yeh and Szu-Wei Fu and Chien-Feng Liao and Elena Rastorgueva and Fran{\c{c}}ois Grondin and William Aris and Hwidong Na and Yan Gao and Renato De Mori and Yoshua Bengio},
  title   = {{SpeechBrain}: A General-Purpose Speech Toolkit},
  journal = {arXiv preprint arXiv:2106.04624},
  year    = {2021}
}

@article{redimnet2,
  author        = {Ivan Yakovlev and Anton Okhotnikov},
  title         = {{ReDimNet2}: Scaling Speaker Verification via Time-Pooled Dimension Reshaping},
  journal       = {arXiv preprint arXiv:2603.11841},
  year          = {2026},
  eprint        = {2603.11841},
  archivePrefix = {arXiv}
}

@inproceedings{ioffe2006plda,
  author    = {Sergey Ioffe},
  title     = {Probabilistic Linear Discriminant Analysis},
  booktitle = {Proc. ECCV},
  pages     = {531--542},
  year      = {2006},
  doi       = {10.1007/11744085_41}
}

@article{auckenthaler2000score_norm,
  author  = {Roland Auckenthaler and Michael Carey and Harvey Lloyd-Thomas},
  title   = {Score Normalization for Text-Independent Speaker Verification Systems},
  journal = {Digital Signal Processing},
  volume  = {10},
  number  = {1--3},
  pages   = {42--54},
  year    = {2000},
  doi     = {10.1006/dspr.1999.0360}
}

@article{brummer2006calibration,
  author  = {Niko Br{\"u}mmer and Johan A. du Preez},
  title   = {Application-Independent Evaluation of Speaker Detection},
  journal = {Computer Speech \& Language},
  volume  = {20},
  number  = {2--3},
  pages   = {230--275},
  year    = {2006},
  doi     = {10.1016/j.csl.2005.08.001}
}

\end{document}